# A Novel Approach to Evaluating Battery Charger Controller Design with Nonlinear PID Controller in an Extendable CHIL Setup

Shervin Salehi Rad
Department of Electrical and Computer Engineerng
Drexel University
Philadelphia, PA, USA
ss5485@drexel.edu

Micheal Muhlbaier
Vice President Of Engineering
Alencon Systme LLC
Philadelphia, PA, USA
mmuhlbaier@alenconsystems.com

Oleg Fishman
Chairman of the Board of Directors
Alencon Systme LLC
Philadelphia, PA, USA
ofishman@alenconsystems.com

Javad Chevinly
Department of Electrical and Computer Engineerng
Drexel University
Philadelphia, PA, USA
js5455@drexel.edu

Elias Nadi
Department of Electrical and Computer Engineerng
Rowan University
Glassboro, NJ, USA
nadiel37@students.rowan.edu

Hua Zhang
Department of Electrical and Computer Engineerng
Rowan University
Glassboro, NJ, USA
zhang@rowan.edu

Fei Lu
Department of Electrical and Computer Engineerng
Drexel University
Philadelphia, PA, USA
fei.Lu@drexel.ed

***Abstract*—The design and development of power electronics converters pose a multitude of challenges. The evaluation of power electronics converters, particularly when operating at high power levels, presents a significant task, offering designers a deeper understanding of the functionality. Several methodologies have been devised to conduct hardware-in-the-loop (HIL) tests, which are classified into two main categories: controller hardware-in-the-loop (CHIL) and power hardware-in-the-loop (PHIL) tests. This paper explores the advantages and drawbacks of these two approaches and introduces a straightforward and cost-effective CHIL method for initial proof of concept and risk-free controller training. This method is based on the interaction between MATLAB/Simulink and Python. To assess the operation of the proposed system, a modeled battery pack (96S1P) is charged using a modeled battery charger converter, employing a non-linear PID controller. In this scenario, the controller benefits from the CC-CV technique for charging the battery pack. The results are presented in the final section.**

***Keywords—Hardware-in-the-loop, Battery charger, Controller hardware-in-the-loop, Power hardware-in-the-loop.***

## I. INTRODUCTION

As the integration of renewable energy sources into power grid systems becomes increasingly prevalent, it becomes imperative to assess the stability and reliability of the employed power electronics converters. It is essential to understand their dynamic responses in various scenarios before deploying them in real-world applications. Hard-ware-in-the-Loop (HIL) testing is a widely embraced solution for evaluation. This method provides designers with results that closely approximate real-world performance and fosters a more detailed understanding of the developed systems. HIL simulation can be categorized into two types: power hardware-in-the-loop (PHIL) and controller hardware-in-the-loop (CHIL) tests [1]-[3].

PHIL involves the integration of physical power components from a converter with a simulated control system. In PHIL tests, the real power hardware is interconnected with a control system simulation, enabling the evaluation of how the control system interacts with real hardware. This approach facilitates a more authentic testing environment by incorporating actual power components, thereby aiding in the detection of issues that may not be evident in a purely simulated setting. It serves as a valuable platform for validating the performance of control strategies. Nevertheless, PHIL setups tend to be more intricate and costly due to the necessity for actual power components and the associated safety measures. Fig 1 provides an overview of the PHIL process [2],[4].

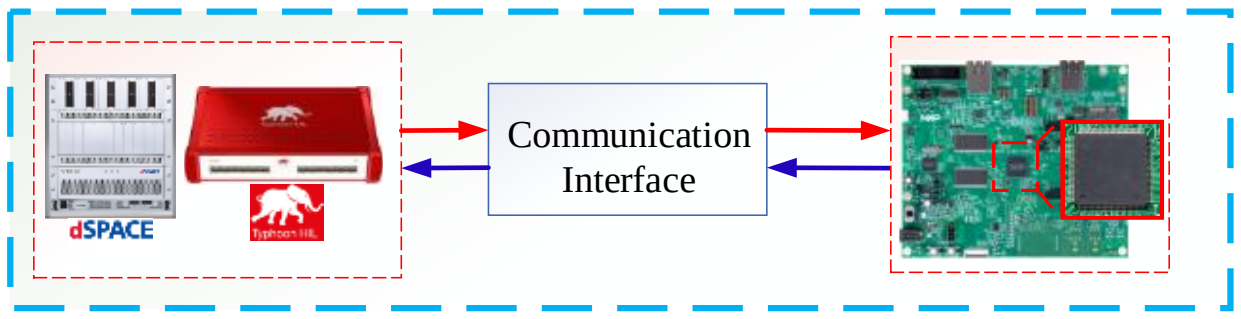

**Fig 1.** Structure of PHIL test platform [6],[7].

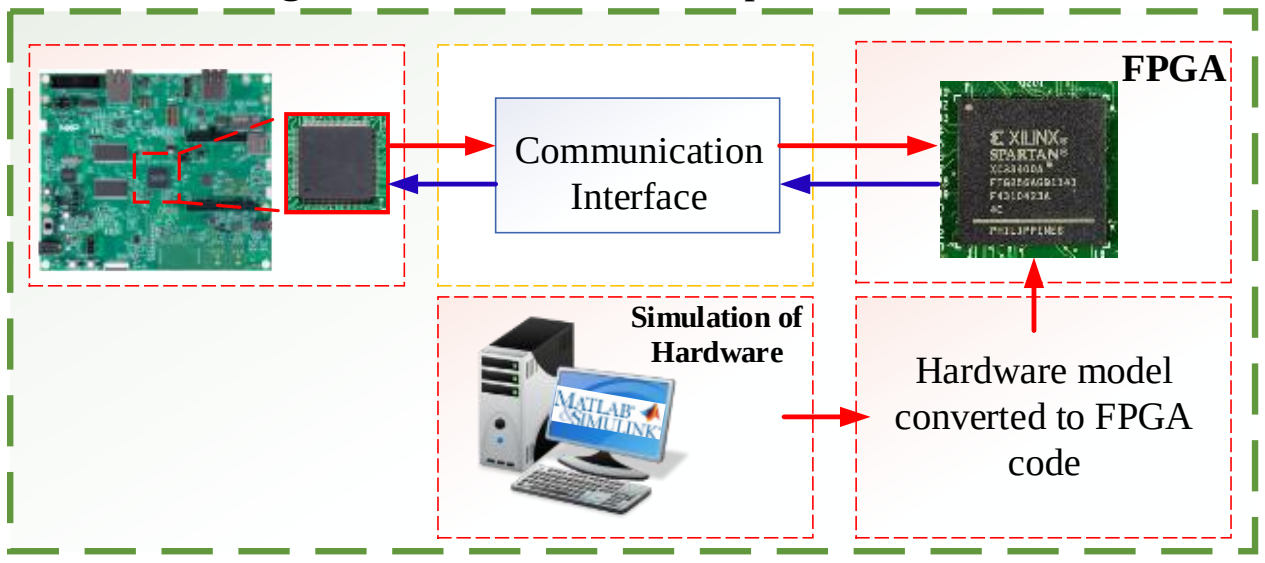

**Fig 2.** Structure of the studied CHIL test platform in this paper.

On the other hand, CHIL testing pertains to the assessment of a power electronics converter's control system within a virtual setting. The physical power hardware is typically emulated, while the control hardware interfaces with the simulation. This approach enables the evaluation of the control system's performance without the necessity of actual power components. It effectively obviates the need for costly physical power hardware, rendering it a more economical choice for preliminary testing and development. There is no risk of damaging real components during testing, as the power hardware is simulated. Nevertheless, it may not perfectly replicate the real-world behavior of the hardware. Fig 2 provides an overview of the PHIL process [5]-[7].

**TABLE 1.** ADVANTAGES AND DISADVANTAGES OF HIL TECHNIQUES

| HIL technique | Advantages | Disadvantages |
|---|---|---|
| PHIL | ○ Realistic Testing<br>○ Real-World Validation | ○ Complexity<br>○ Safety Concerns |
| CHIL | ○ Cost-Effective<br>○ Risk-free<br>○ Flexibility | ○ Model Accuracy |

Table 1 compares HIL techniques. The selection between CHIL and PHIL depends on the precise testing requisites and the developmental stage. Typically, CHIL is the preferred choice for preliminary testing and controller validation, given its cost-efficiency and safety advantages. On the other hand, PHIL becomes prominent when the control strategy necessitates validation under realistic, high-power conditions, despite its increased complexity and cost [2]. Moreover, since testing the novel and designed techniques would be a serious challenge for power electronics converters in terms of safety and reliability especially in high-power applications, there is an urgent need for a platform to assist engineers in resolving this issue.

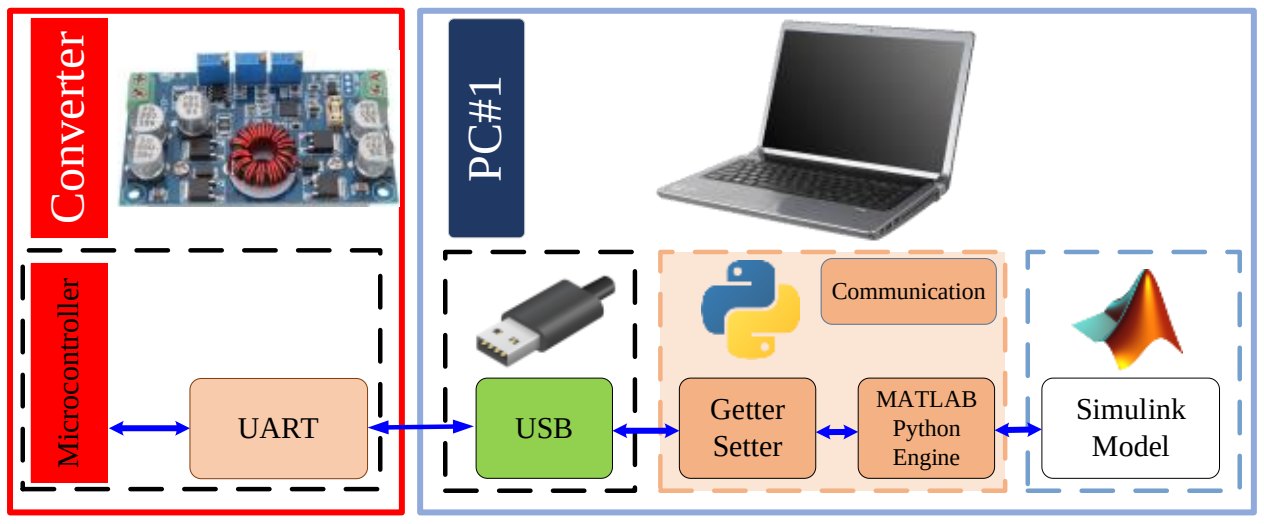

**Fig 3**. Proposed CHIL simulation technique [8]

## II. PROPOSED CHIL TECHNIQUE

This paper introduces a novel CHIL setup aimed at simplifying the system and achieving cost-effective implementation. It removes the requirement for high-speed microcontrollers (MCUs) and FPGAs for translating the hardware system into code. Instead, it operates within the MATLAB/Simulink environment. An additional benefit lies in its event-based sampling capability. Furthermore, it offers the advantage of facilitating the connection of multiple MCUs without inherent limitations, thus enabling system expansion. This approach involves the utilization of a MATLAB engine API for Python, serving as the intermediary communication interface between MATLAB/Simulink and MCUs. The system's configuration is visually depicted in Fig 3.

The PC#1 section is a decisive factor in CHIL testing, this section divides into three parts, MATLAB, communication, and USB sections. In MATLAB, the hardware model of the converter is drawn and run. MATLAB benefits from an extension called “MATLAB Python Engine” that can extract the defined measurements from a simulation and save it; meanwhile, getter_setter unit is designed and developed in PC#1 to get data from MATLAB and put them in its parameter to send it to the MATLAB python engine. Moreover, the getter_setter section is used to receive data that is calculated by the MCU and send it to the MATLAB python engine to be sent to MATLAB model. It has a high degree of freedom in defining parameters. Some parameters are shown in Table 2.

**TABLE 2.** MATLAB AND MICROPROCESSOR DATA

| MATLAB data to MCU | | MCU data to MATLAB | |
|---|---|---|---|
| # | Parameter | # | Parameter |
| 1 | Primary voltage | 1 | Ready (Enable) |
| 2 | Secondary voltage | 2 | Running mode |
| 3 | Primary current | 3 | Time of operation |
| 4 | Secondary current | | |

The getter_setter unit, which has been designed, consistently remains in a state of readiness, awaiting data transmission from MATLAB in PC#1. Once data is received, the getter_setter unit facilitates data exchange. The getter_setter unit also encompasses the task of transmitting data, which is received through USB, to the MATLAB/Simulink section. In situations where there is no data available for transmission to either MATLAB or the MCU, the getter_setter unit switches from 'ready' mode to 'pause' mode, thereby notifying both devices. This behavior is shown in Fig 4.

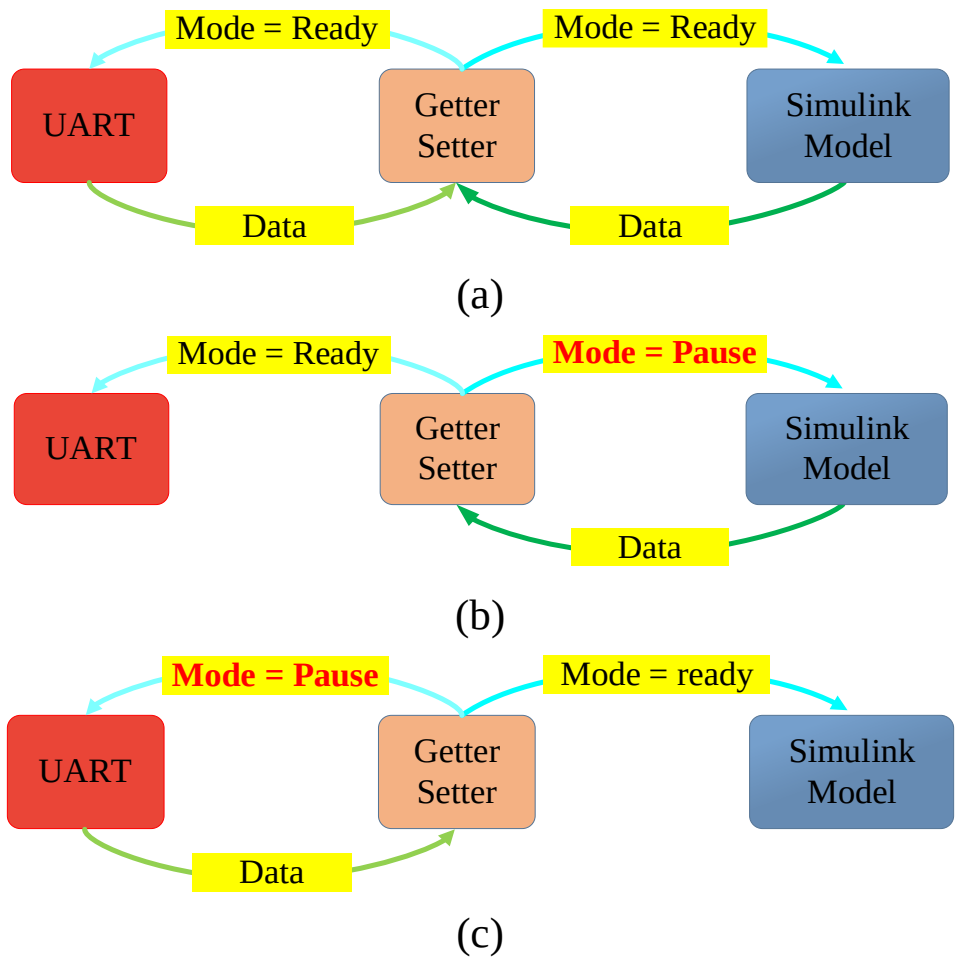

**Fig 4**. MATLAB and microprocessor data exchange. a) standard mode, b) microprocessor requires additional time, and c) MATLAB requires extra time.

## III. Non-Linear PID Controller Technique

One of the challenging parts in controlling power electronics converters is PID controller. This unit is utilized to improve the system's response and performance. There are numerous techniques to design and implement PID controllers. Non-linear and adaptive PID controllers are the most popular controllers in this field [9].

The main difference between non-linear PID controllers and adaptive PID controllers lies in their approach to handling system dynamics and uncertainties. Non-linear PID controllers focus on incorporating non-linearities directly into the control law, while adaptive PID controllers dynamically adjust their parameters to adapt to changes in system behavior or uncertainties. Both types of controllers can offer improved performance in challenging control system applications, but they address different aspects of control system design and operation [9]-[10].

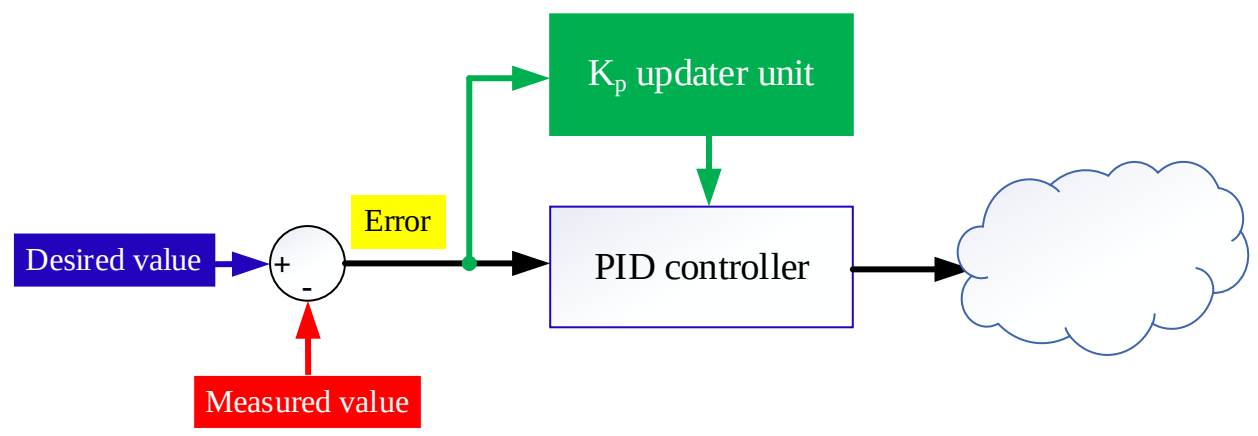

**Fig 5**. Structure of the implemented non-linear PID controller.

A Nonlinear PID controller is an enhanced version of the classic PID controller used in control systems [9]. The nonlinear PID controller incorporates nonlinear elements to handle systems with complex and non-linear behaviors [11]. Since the battery charger converters have non-linear dynamics, in this paper, a non-linear PID controller is deployed on an MCU. Its parameters adapt based on the difference between reference and measured voltages or currents. Fig 5 illustrates the concept behind this method.

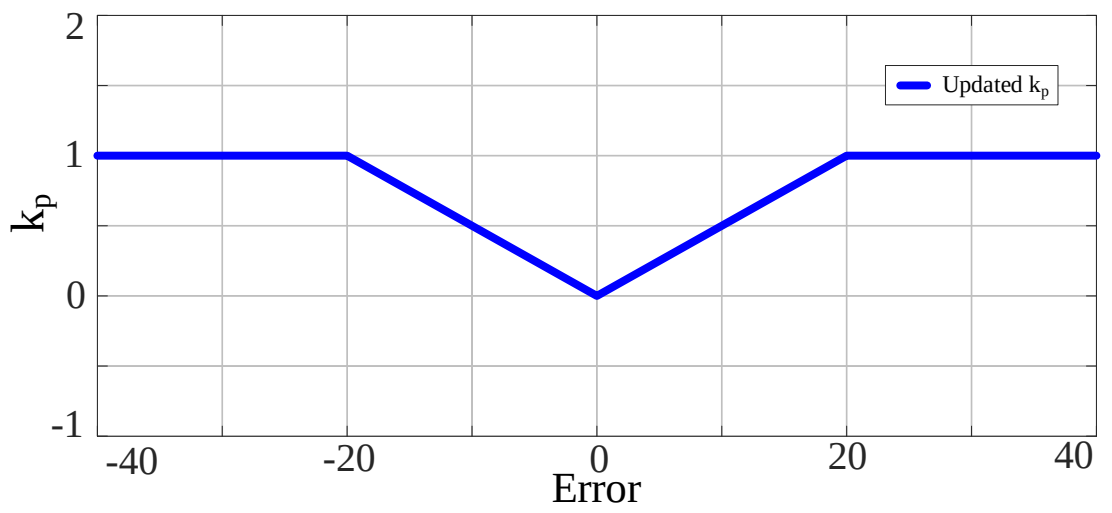

**Fig 6**. Updating $k_p$ parameter in non-linear PID controller

To clarify the developed model of the non-linear PID controller, the behavior of the designed controller for only the parameter $k_p$ is shown in Fig. 6. According to the system model, the allowable range for $k_p$ should be between 0 and 1. When the difference between the desired value and the measured value exceeds 20 (error is greater than or equal to 20), $k_p$ is set to 1 to signal the PWM unit to maximize its duty cycle to reach the desired value. Conversely, when the difference between the desired value and the measured value is approximately 10, $k_p$ is updated to 0.5 to signal the PWM unit to maintain awareness of this situation and prevent an excessive increase in the duty cycle to reach the desired value.

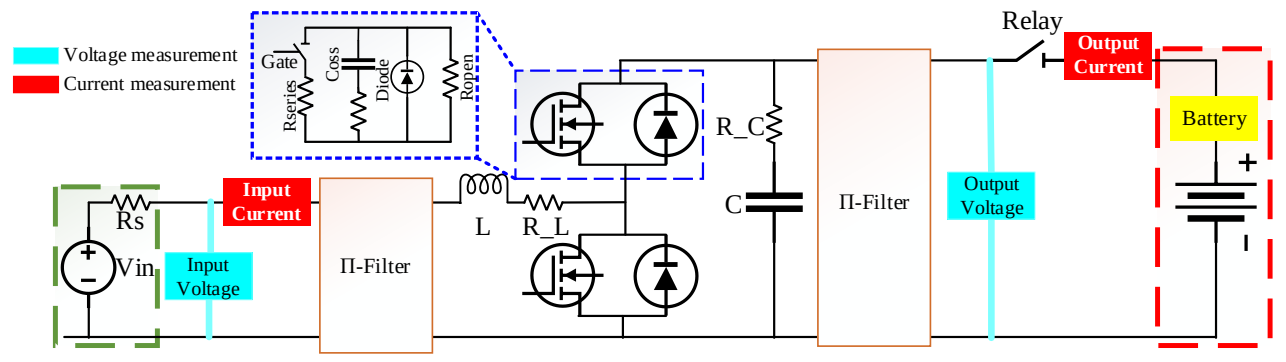

**Fig 6.** Schematic of al battery charger implemented in MATLAB/Simulink

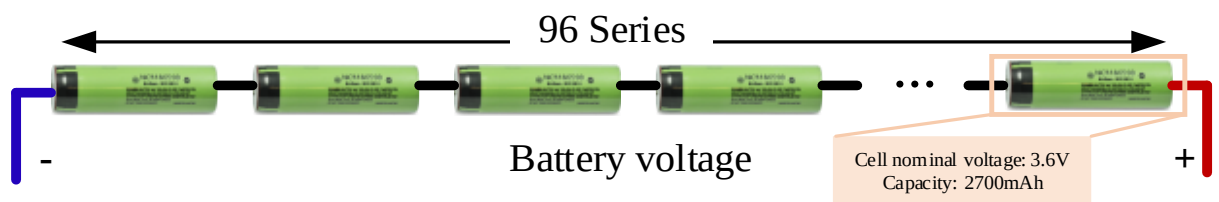

**Fig 7**. Modeling of the designed batteries.

It is important to note that the negative error in Fig. 6 occurs when the overshoot happens. In the next section, the designed non-linear PID controller will also be implemented and uploaded to the MCU to test the system's response and performance.

## IV. Experimental CHIL Results

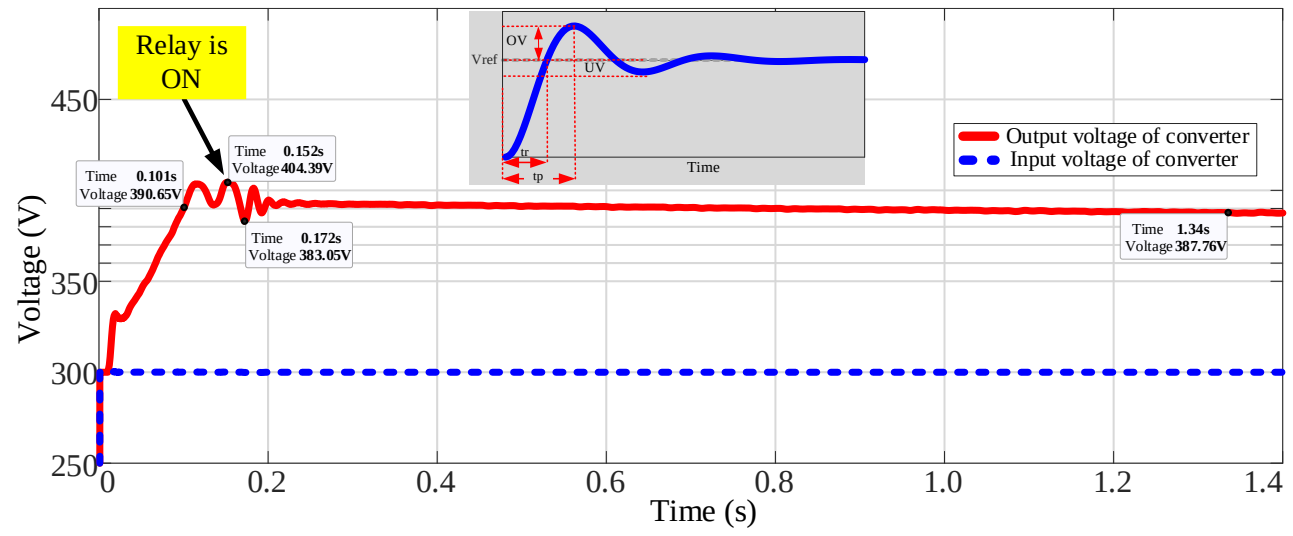

**Fig 8.** Voltage at both the Input and Output of the Boost Converter during CC mode.

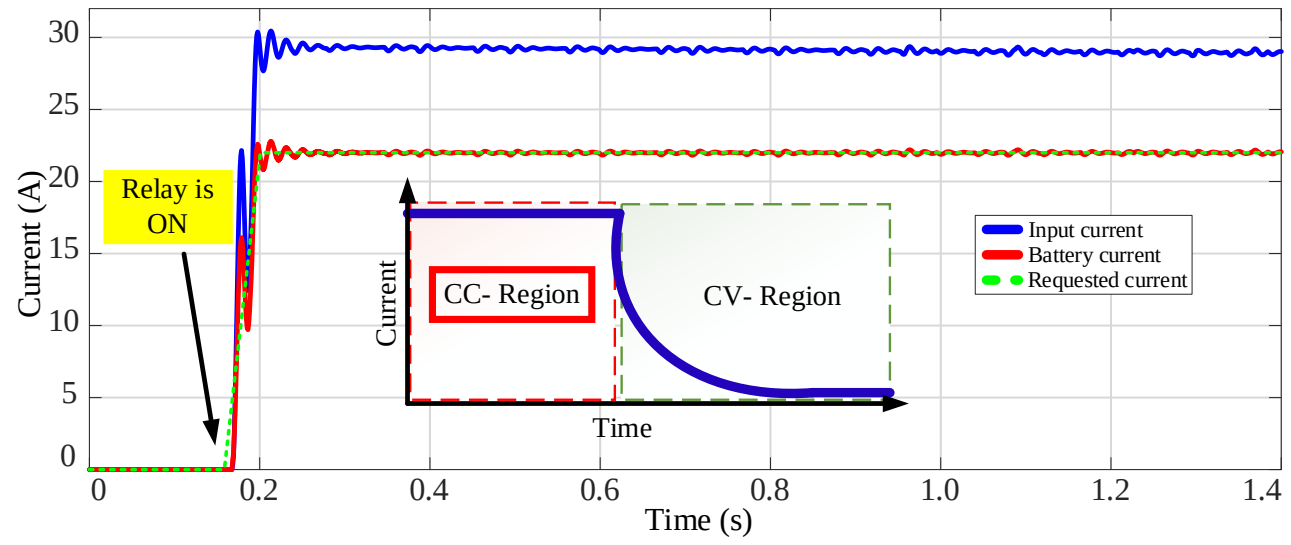

**Fig 9.** Current at the Input and Output of the Converter during CC mode.

A battery charger undergoes testing using the proposed CHIL platform. To assess the functionality of the implemented code and detect any potential unforeseen errors, a MATLAB/Simulink model of the battery charger is constructed. The system model, depicted in Fig. 6, includes an input voltage rated at 300V and an internal resistance of 250 milliohms. It comprises a Boost converter to convert the 300V input to 400V, a relay for protection against inrush current, and a battery model. A model of the NCR18650 lithium-ion battery cell is utilized in MATLAB Simulink. The designed code has been uploaded to the MKV58F1M0VLL24 MCU from NXP company, with the specifications of both the MCU and battery detailed in Table 3. Since the objective of this system is to replicate an EV battery charger, 96 of these cells, as depicted in Fig. 7, are connected in series to raise the voltage to 400V.

**TABLE 3.** MCU AND BATTERY SPECIFICATIONS

| Parameter | Value |
|---|---|
| **MKV58F1M0VLL24** | |
| MCU operating frequency | 220MHz (Real-time control MCU) |
| **NCR18650** | |
| $C_{nominal}$ | 2700 mAh |
| $V_{max}$ | 4.2V |
| $V_{cutoff}$ | 3.0V |
| $V_{nominal}$ | 3.6V |

"The state of charge (SoC) for the entire system is set to 50%, and the converter initiates the battery charging process with a high current of 23A. The MCU is programmed to manage the battery charging based on a constant current-constant voltage (CC-CV) algorithm.

The test is divided into two phases: the CC test and the CV test. In the initial test, the battery is charged by the system functioning in CC mode. Subsequently, in the next test, the system's behavior is observed using the CV technique. The voltage levels at the input and output terminals of the converter are illustrated in Fig. 8.

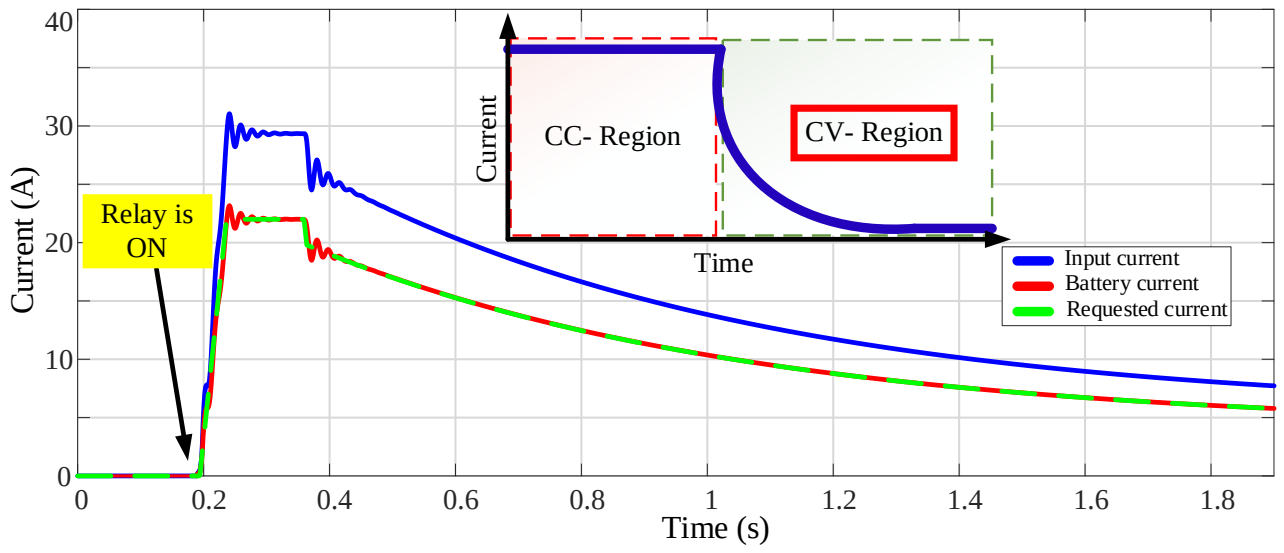

**Fig 10.** Current levels at both the Converter's input and output, along with the intended current from the MATLAB side during CV mode

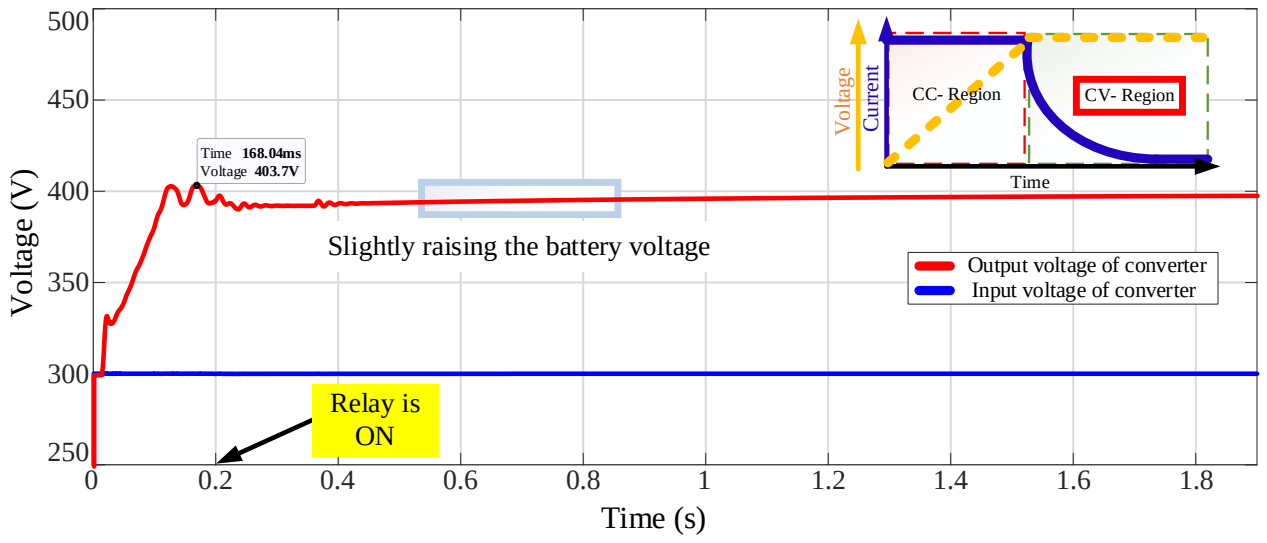

**Fig 11.** Voltage at the input and output of the Boost Converter in CV mode

The system and filters are stabilized in less than 200ms. Subsequently, as shown in Fig. 6, the relay is connected to the battery side to start the charging process. The primary objective in this scenario is to maintain a constant current at the battery side for charging, as shown in Fig. 9.

As observed, the battery charger exhibits a minor overshoot of 1.1% in the operational mode (with a reference voltage of 400V and a maximum voltage of 404.39V). Once connected to the load, it stabilizes in under 20ms.

In the following test, aimed at assessing the system's performance in CV mode, the battery's SoC is adjusted to 90%. The system is then operated in CV mode, where an initial injection of 23A is provided. Subsequently, the controller monitors the system's behavior, and the current automatically decreases to facilitate battery charging. Fig. 10 and Fig. 11 provide the CHIL simulation results of the battery's current and voltage waveforms, respectively. As can be seen, thanks to the well-designed non-linear PID controller parameters, the voltage overshoot is only 0.925% in this situation (with a reference voltage of 400V and a measured value of 403.7V).

## V. CONCLUSIONS

Power electronics converters face a significant challenge in implementing designed controllers and testing them in real-world scenarios. Testing new algorithms is often difficult, especially when it involves critical situations that can be complex and hazardous. To address this issue, this paper explores hardware-in-the-loop (HIL) techniques, which offer designers an easier way to test their algorithms before deployment. However, traditional HIL techniques are often costly and require complex platforms. In response, this paper introduces a novel, cost-effective, and safe approach based on MATLAB/Simulink and Python code. This technique is not only easy to extend but also compatible with all types of microcontrollers (MCUs), providing designers with greater flexibility. The paper elaborates on the operation of the proposed CHIL technique. To evaluate its performance, a conventional battery charger is used to charge a modeled battery pack (96S1P) employing the CC/CV method. Additionally, the paper introduces a straightforward non-linear PID controller to enhance the system's response. Finally, the paper presents all CHIL experimental results.

## VI. ACKNOWLEDGEMENTS

The research team would like to express appreciation for the support received from Alencon Systems LLC, supervised by Dr. Oleg Fishman and Dr. Michael Muhlbaier. The views and opinions expressed herein by the authors do not necessarily reflect those of the United States Government or any of its agencies.

We would also like to acknowledge the valuable discussions with Dr. Michael Muhlbaier, Vice President of Engineering at Alencon Systems, and Dr. Oleg Fishman, Chairman of the Board of Alencon Systems LLC.

This project has been sponsored by Alencon Systems LLC; however, the company neither endorses nor rejects the findings of this research.